# Scope of cloud computing for SMEs in India

Monika Sharma, Ashwani Mehra, Haresh Jola, Anand Kumar
Dr.Madhvendra Misra, Ms.Vijayshri Tiwari
Indian Institute of Information Technology Allahabad (IIITA), Allahabad, India

*Abstract*— Cloud computing is a set of services that provide infrastructure resources using internet media and data storage on a third party server. SMEs are said to be the lifeblood of any vibrant economy. They are known to be the silent drivers of a nation's economy. SMEs of India are one of the most aggressive adopters of ERP Packages. Most of the Indian SMEs have adopted the traditional ERP Systems and have incurred a heavy cost while implementing these systems. This paper presents the cost savings and reduction in the level of difficulty in adopting a cloud computing Service (CCS) enabled ERP system. For the study, IT people from 30 North Indian SMEs were interviewed. In the cloud computing environment the SMEs will not have to own the infrastructure so they can abstain from any capital expenditure and instead they can utilize the resources as a service and pay as per their usage. We consider the results of the paper to be supportive to our proposed research concept.

## 1. INTRODUCTION

Today is the age of information technology. The facets of work and personal life are moving towards the concept of availability of everything online. Understanding this trend, the big and giant web based companies like Google, Amazon, Salesforce.com came with a model named "Cloud Computing " the sharing of web infrastructure to deal with the internet data storage, scalability and computation (Kambil,2009). According to the definition by NIST "Cloud computing is a model for on-demand network access to a shared pool of configurable computing resources that can be rapidly provisioned and released with minimal management effort or service provider interaction".

Cloud computing is an online service model by which hardware and software services are delivered to customers depending upon their requirements and pay as an operating expense without incurring high cost (Bandyopadhyay et al, 2009). Basically cloud computing is a set of services that provide Infrastructure resources using Internet media and data storage on a third party server. It has three dimensions known as Software level service, Platform level service, Infrastructure service (Fox, 2009).

The main cloud computing attributes are pay per use, elastic self provisioning through software, simple scalable services, virtualized physical resources (Tucker). Models, such as cloud computing based on Virtual technologies enables the user to access storage resources and charge according to the resources access (Marcos et al, 2009). Cloud computing platforms are based on utility model that enhances the reliability, scalability, performance and need based configurability and all these capabilities are provided at relatively low costs as compared to the dedicated infrastructures (Wyld, 2009). Benefits provided by cloud computing ranges from cost savings to speed and flexibility to enhanced performance(Veverka,2010).This new model of infrastructure sharing is being widely adopted by the industries (Hartig,2008). Industries experts predicts that cloud Computing has bright future in spite of changing technology that faces significant challenge (Leavitt, 2009). The report from IDC says that due to the emergence of cloud computing, IT marketplace is undergoing a change and it expects that investment on cloud services will reach to $42 billion by 2012.

## 2. CLOUD COMPUTING IN INDIA

India is growing at faster pace in information technology sector thereby showing a great potential for the cloud computing services. According to Springboard Research report (Jan 2009) SAAS India i.e. software as a service in India will register a compounded annual growth rate of 76% in the time period of 2007-2011. Cloud computing services has huge opportunity in Indian market due to the large number of Small and Medium businesses (SMBs) which is at around 35 million and they want easy to use , reliable and scalable application that helps them to grow and expand their business. This makes India as the fastest growing SAAS market in Asia Pacific region. According to Jeremy Cooper, VP-Marketing (APAC), Salesforce.com 'software as a service' provider started its services in India in September 2005 and since then the adoption rate of cloud computing is increasing .Seeing SAAS success on September 2008 IBM launched cloud computing center in India at Bangalore. This center will cater to the increasing demand of web based infrastructure sharing services. IBM India collaborated with IIT Kanpur to come up with some new developments in computing that will help in academic advancement. Bharti Airtel has launched the cloud computing services with their NetPc model and other giant companies like Reliance Communications, TCS, HCL technologies, Wipro, Netmagic, Verizon, Novatium etc. have also launched cloud computing services in India.

### 2.1 Cloud computing For the SMEs in India

SMEs are said to be the lifeblood of any vibrant economy. They are known to be the silent drivers of a nation's economy. SMEs are leading the way for entering new global markets and for innovations in the emerging economic order. In India 95% of the industrial units are SMEs which give over 50% of the industrial output (Popli and Rao, 2009).Thus SMEs form the backbone of the Indian economy.

SMEs of India are one of the most aggressive adopters of ERP Packages. Online services are better suited for small industries whereas large enterprises face more problems in implementation because of their complex functionalities and data security concerns (Dubey and Wagle, 2007). Small and Medium businesses have sufficient IT budgets to buy the bandwidth and pay as per their need and usage. The main problem faced by the S.M.E.'S when it comes to traditional ERP implementation is that their requirements are limited while the product offered always exceeds their specifications in every way (including the costs). This gap between the SMEs Requirements and the traditional ERP's specifications needs to be analyzed by the companies (traditional ERP providers) and the SMEs. It is not possible for the traditional ERP providers to bring down their standards for the sake of the S.M.E.'s neither is it feasible for the later to upgrade for the sake of the former. Either of these if done leads to direct monetary losses for either one party or both.

According to a paper published by the Associated Chambers of Commerce and Industry of India SMEs sector is growing at a rate of 35% per annum and it will increase to 40% in the coming years (Assocham, 2009). As per the ASSOCHAM report, 60% of SMEs are moving towards the technology based infrastructure to increase their productivity with the reduction in their input cost. SMEs are one of the growing sectors and lucrative market places for the implementation of enterprise solutions. As the traditional in-house implementation of ERP solutions incurs high cost for the SMEs so it becomes a major constraint for them. So our purposed cloud computing cost efficient model is based on leveraging the cloud web services as a substitute for ERP solution by paying only for what the SMEs actually use. By using and accessing services through the cloud, the companies can buy components relevant to their business on pay per basis instead of buying whole ERP suite. (Sharif, 2009).

In the cloud computing environment the SMEs will not have to own the infrastructure so they can abstain from any capital expenditure and instead they can utilize the resources as a service and pay as per their usage of the resources provided by the cloud (Rittinghouse and Ransome , 2009). SaaS will provide an opportunity for the SMBs to automate their business by reducing their investment in IT infrastructure (Rao).Cloud based services helps the industries to reduce their cost that are involved in on-premise ERP solutions such as hardware, software, upgradation, training and licensing costs. Moreover long implementation cycles with regular maintenance costs adds to the total cost of traditional ERP (Aggarwal and Barnes, 2010).

According to V Ramaswamy, SMB global head (TCS), SMBs are in need of easy to use technology (Business standard, Jan 2010). With the changing needs and increase of customer base there is requirement of CRM and ERP solutions. As technology changes companies requires up gradation in their software this poses obstacles for the SMEs to scale up. In order to operate in limited budget a less complicated and simplified offering is required. At present most of the Indian and Foreign IT companies are focusing on SMEs for their cloud computing offerings. According to Leslie D'Monte, Cloud computing is providing huge opportunities for the Indian IT company that is helping them to develop cost effective business models. Such models help the SMEs to uplift their business in an effective and cost efficient manner. The promoter of 'The India Cloud Initiative' Vijay Mukhi said that there is a huge saving of money by using cloud technology as the industries have to pay only for the operating cost. The biggest advantage of a hosted model (cloud computing) is that it eradicates the need to purchase the software licenses and also eliminates the cost associated with developing and operating in-house applications. In a hosted model, the capital investment, security, backup and server maintenance costs are all the provider's responsibilities."

### 3. HYPOTHESIS DEVELOPMENT

Cloud Computing is web based subscription model enabling the users to pay as per their need and usage. Cloud Computing Model provides IT based services and capabilities online with data shared on a third party server. As the users are paying on hourly basis and in some cases on monthly basis, cloud computing will result in a substantial cost saving and it will leverage the benefits of ERP solutions. Hence we hypothesize:-

H1 : Cloud computing service provide lower per user annual cost than traditional ERP system.

H2 : Cloud computing service provide higher per user annual cost than traditional ERP system.

H3 : Cloud computing services are more adaptable than traditional ERP systems.

H4 : Cloud computing services are less adaptable than traditional ERP systems.

### 4. RESEARCH METHODOLOGY

This paper is a study of the scope of cloud computing for SMEs in India. The Main purpose of this paper is to examine and analyze the Scope of cloud computing for the SMEs in India. So, this research paper aims to develop a research model which would justify this papers affinity towards the use of cloud computing for Indian SMEs. One of the major challenges was to get the financial data of some SMEs and another was to understand which data to choose for analyzing the scope.

This research paper adopts a "DESCRIPTIVE TYPE" of research. The research methodology used for the paper was kept very simple. Primary data was collected by conducting Telephonic interviews and by questionnaires sent through email and Secondary data was collected from the internet, news papers, magazines and journals. T- Test was applied on the collected data to test one of the hypothesis of this paper that using cloud computing based ERP software would cost the SMEs lesser than the traditional ERP software does. Factor rating method was used to test the other hypothesis that traditional ERP systems involve higher level of difficulty in terms of adaptability than the Cloud computing services.

### 3.1 Data Collection Methods:

1. Primary data: The Primary data for this paper was collected using telephonic interview with IT personnel from about 30 Indian SMEs. It was a Structured Interview which comprised of sequenced questions.

   An E-mail based questionnaire was also used to collect the primary data. This questionnaire was sent to users of various traditional ERP software's in the above 30 Indian SMEs to get the general user opinion used to test the hypothesis $H_3$.

2. Secondary data: The Secondary data was collected using
   a) Internet: The Sites of all ERP providing software's using SaaS technology were visited and the cost per user for each of them was collected via this method. The SAP ByDesign Software's site was well studied and even non numbered data was collected from the site.
   b) Journals: White papers from emerald Journal were used to collect data.
   c) News Papers: Business Standard daily business news paper was also used as a source for secondary data collection.
   d) Magazines: The CTO Forum magazine, Business Week.

## 5. ANALYSIS AND INTERPRETATION

Starting from the above mentioned information that SMEs will incur lower cost by using cloud computing services than by using traditional on premise ERP systems, the hypothesis $H_1$ and $H_2$ are analyzed as:

In order to get the information about the ERP implementation cost, those SMEs were targeted where the IT investment cost ranges from 65 to 90 lakhs. The data that was collected was about per user per year cost incurred by the SMEs and the cloud computing prices offered by the different companies for the same units. This is illustrated in Table 1 and Table 2 respectively.

Table 1:-

| NAME OF THE COMPANY | Software Used | Avg. Software Cost | Avg. Hardware Cost | Avg. Connectivity Cost | Total cost | No. of users | Cost /User |
|---|---|---|---|---|---|---|---|
| \multicolumn{8}{|c|}{ERP UTILISATION REPORT (Amount Rs. in Lacs)} |
| Rana Sugars Ltd. | ORACAL FINANCE | 40.00 | 20.00 | 15.00 | 75.00 | 61.00 | 1.22 |
| Sigma Vibracoustic (I) Pvt.Ltd | SAP | 42.00 | 21.00 | 16.00 | 79.00 | 75.00 | 1.05 |
| Pritika Industries Pvt.Ltd | RAMCO | 38.00 | 20.00 | 14.00 | 72.00 | 50.00 | 1.44 |
| Hind Motors Ltd | PEOPLE SOFT | 35.00 | 18.00 | 14.00 | 67.00 | 52.00 | 1.28 |
| Cheema Spinnings Ltd | NAVISION | 45.00 | 25.00 | 17.00 | 87.00 | 65.00 | 1.33 |
| Kudos Chemie Ltd. | ORACAL FINANCE | 42.00 | 25.00 | 18.00 | 85.00 | 57.00 | 1.49 |
| Nector Pharma Pvt.Ltd | PEOPLE SOFT | 38.00 | 20.00 | 12.00 | 70.00 | 52.00 | 1.34 |
| Winsome Textiles Ltd. | ORACAL FINANCE | 40.00 | 22.00 | 17.00 | 79.00 | 65.00 | 1.21 |
| Jaypee Coach Builders Ltd | BAAN | 35.00 | 17.00 | 15.00 | 67.00 | 58.00 | 1.15 |
| Teva Pharmaceuticals Ltd. | PRIORITY | 42.00 | 22.00 | 14.00 | 78.00 | 65.00 | 1.20 |

Table 2:-

| SAP Business ByDesign ($149 per user per month) | Salesforce Enterprise ($125/user/month) | Professional ($65/user/month) | Force.com unlimited ($75/user/month) |
|---|---|---|---|
| 0.822 | 0.69 | 0.358 | 0.414 |

After analyzing the different cost structures of various companies offering cloud computing services it was found that charges were less than Rs.83000 per user per year approximately. Depending upon this information, t-test was applied to know whether using cloud computing has low or high cost than on-premise ERP cost. On applying the test we get the following result:

| Hypothesis | $H_1$ |
|---|---|
| Mean | 1.22 |
| S.D. | 0.214 |
| T-Test | 0.588 |
| DF | 9 |
| R:t | >1.833 |
| SL | 5% |
| Result | Accepted |

To determine the rejection region or acceptance region, one tailed test is applied at 5 percent level of significance using table of t-distribution for 9 degree of freedom we get the value:

R: t>1.833

The observed value of t is 0.588 which is in the acceptance region and thus H1 is accepted at 5 percent level of significance and it can be concluded that the sample data indicates that traditional on-premise ERP incurred higher cost as compared to cloud computing services.

### 5.1 Graphical analysis

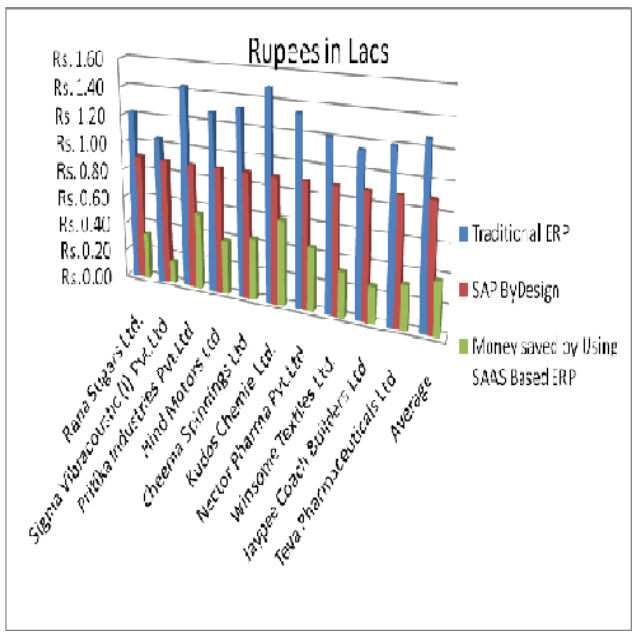

Figure 1

The above Graph compares the cost which has been incurred by different companies for using traditional ERP software. Alongside this cost the second bar gives the cost which the companies would have incurred if they would have used SAP ByDesign, the SaaS based ERP solution. The Third bar indicates the amount of money that would have been saved by these companies if they would have used SAP ByDesign, the SaaS based ERP solution. The y-axis indicates the money in lakh Rupees and the x-axis presents the name of the companies using the traditional ERP systems. It is clearly evident from the graph that most of these companies could have on an average saved more than Rs. 37000 per user per year by using SAP ByDesign the SaaS based ERP instead of the traditional ERP.

According to a study conducted by the research firm Gartner in the year 2008, adoption of hosted applications like SAP Business ByDesign, reduces cost of ownership by about 30% by lowering the software support, labor and hardware costs. This study further supports the papers findings.

### 5.2 Analysis using Factor Rating Method

In addition to the cost there are certain more factors based on which decision has to be made while implementing the cloud computing services for SMEs. So hypothesis H3 and H4 are analyzed:

To test this hypothesis factor rating method was used and the factors were analyzed on a scale of 3 in terms of difficulties faced by ERP and cloud computing customers.

1= Low difficulty

2= Moderate difficulty

3= High difficulty

Table below presents the different factors of adoption with their score and weighted score.

| Parameter | ERP Score | Cloud Computing Score | Weights | ERP Weighted Score | Cloud Computing Weighted Score |
|---|---|---|---|---|---|
| Scalability | 3 | 1 | 7.5 | 22.5 | 7.5 |
| Availability | 2 | 1 | 8.5 | 17 | 8.5 |
| Maintainability | 3 | 1 | 7 | 21 | 7 |
| Accessibility | 2 | 1 | 6.5 | 13 | 6.5 |
| Mobility | 3 | 1 | 7.5 | 22.5 | 7.5 |
| Performance | 2 | 1 | 7.5 | 15 | 7.5 |
| Implementation | 3 | 1 | 7.5 | 22.5 | 7.5 |
| Security | 1 | 3 | 8.5 | 8.5 | 25.5 |
| Integration | 1 | 3 | 8 | 8 | 24 |
| Deployment | 3 | 1 | 5 | 15 | 5 |
| Flexibility | 3 | 1 | 5.5 | 16.5 | 5.5 |
| Transparency | 2 | 2 | 6 | 12 | 12 |
| Installation | 3 | 1 | 7.5 | 22.5 | 7.5 |
| Up gradation | 3 | 1 | 7.5 | 22.5 | 7.5 |
| Sum | | | 100 | 238.5 | 139 |

Table 3(Analysis of factors of adoption)

Based on the above factors and their weighted scores the graphical analysis was done:

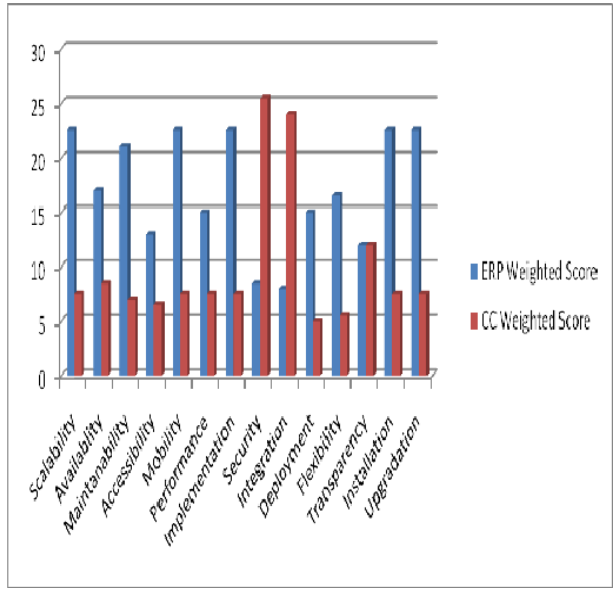

Figure 2

Figure 2 shows that ERP has high weighted score in terms of every factor cited in table 1 except for security and integration. The value of Cloud Computing Weighted Score comes out to be 139 which is less as compared to traditional ERP weighted score as shown below in figure 3. So this analysis of data collected using E-mail based questionnaires from general users indicates that cloud computing has more factors that are less difficult for adaptation than traditional ERP and hence hypothesis 3 is accepted whereas hypothesis 4 is rejected.

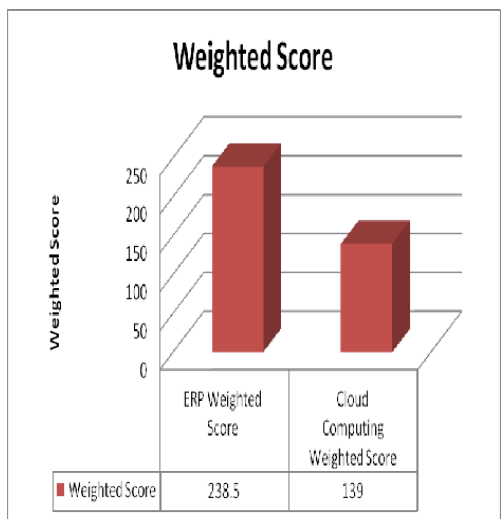

Figure 3

## 6 FINDINGS

From the above analysis based on the results obtained from t-test it is construed that cloud computing incurs lower cost than traditional ERP Systems. No capital investment for software infrastructure is required for any SaaS based ERP solution. So there are no hardware, software or implementation costs, which essentially are responsible for the unprecedented high cost of using a traditional ERP system.

Based on the financial data of the Indian SMEs that are analyzed above it is deduced that on an average these Indian SMEs would have saved approximately Rs. 37000 per user per year if they had preferred the SaaS Solution provided by "SAP Business ByDesign" instead of their already installed traditional ERP systems.

So using the cloud computing model would help the SMEs in optimizing their total cost incurred on ERP by higher per user annual savings.

Based on the factor rating method it can be easily deduced that traditional ERP systems involve higher level of difficulty when analyzed in terms of adaptability than the Cloud computing services. So it is clearly evident that Cloud Computing services are more adaptable than traditional ERP systems.

## 7 CONCLUSIONS

The objective of this research paper was to analyze the scope of cloud computing for the SMEs in India. For this purpose the paper analyzes per user annual cost as a parameter to compare the cost of using the traditional ERP solution and the cloud computing modeled SaaS based ERP systems. The paper also compares the difficulty level for adaptability of the traditional ERP systems and the SaaS based ERP solution. After the analysis following conclusions are drawn:

The average amount saved by using the SaaS based ERP instead of the traditional ERP is about 37000 per user per year for the SMEs under consideration. So our hypothesis (H1) that Cloud computing service provides lower per user annual cost than traditional ERP systems is accepted.

Traditional ERP systems involve higher level of difficulty in terms of adaptability than the Cloud computing services. So our hypothesis (H3) that Cloud Computing services are more adaptable than traditional ERP systems is also accepted.

REFERENCES

[1] S. Bandyopadhyay, S. R. Marston, Z. Li, A.Ghalsasi,"Cloud Computing: The Business Perspective", November 2009.
[2] L. Tucker, "Introduction to Cloud Computing....... for Enterprise Users" Cloud Computing Sun Microsystems, Inc.
[3] L. M. Vaquero, L. R. Merino, J. Caceres, M. Lindner,"A Break in the Clouds: Towards a Cloud Definition", vol. 39, no.1,January 2009.

[4] A. Kambil, "A head in the clouds", vol. 30, no. 4, pp. 58-59, 2009.

[5] R. Fox, "Digital Libraries:The systems analysis perspective", Library in the Clouds, vol. 25, no. 3, pp. 156-161, 2009.

[6] N. Leavitt, "Is cloud computing really ready for prime time?", vol. 42, no.1, pp 15-20, 2009.

[7] D. C. Wyld, "The utility of Cloud Computing as a new pricing-and consumption-model for information technology", Vol. 1 No.1, 2009.

[8] W. Ashford, "Cloud clears the way for SME innovation (cloud computing)", 2008.

[9] K. Hartig, ''What is cloud computing?'', 2008.

[10] A. Dubey and D. Wagle, "Delivering Software as a Service,The McKinsey Quarterly Web Exclusive" 2007.

[11] B.B Aggarwal and M. Barnes, "The Case For Cloud CRM In India" ,White Paper, Springboard Research , 2010.

[12] M.L.N. Rao, "SaaS Opportunity for Indian Channels" Jamcracker Inc.

[13] M. Veverka, "Sky's the Limit", vol. 90 , Iss. 1, pg. 19 , 2010.

[14] A.M. Sharif, "It's written in the cloud: the hype and promise of cloud computing" , Brunel Business School, Brunel University, Uxbridge, UK, 2009.

[15] M. Dias de Assunção, A. Di Costanzo, R. Buyya, "Evaluating the Cost-Benefit of Using Cloud Computing to Extend the Capacity of Clusters", 2009.

[16] J.W Rittinghouse and J.F. Ransome, "cloud computing implementation, management and security", CRC press Taylor & Francis Group 2009.

[17] G.S Popli and D.N. Rao ,"An Empirical Study of SMEs in Electronics Industry in India: Retrospect & Prospects in post WTO Era" 2009.

[18] S. Yang and K. Wang, "The Influence of Information Sensitivity Compensation on Privacy Concern and Behavioral Intention",vol. 40, 2009

[19] P.Godbole, 'Only 10 per cent of SMBs exploit the power of IT',*The Business Standard* ,available at:

http://www.business-standard.com/india/news/%5Conly-10-per-centsmbs-exploitpowerit%5C/381627/( accessed 19 Jan 2010).

[20] Cloud Security Alliance ,"Security Guidance for Critical Areas of Focus in Cloud Computing" 2009.

[21] Nasscom Blog ,'India market and the SaaS/Cloud Computing landscape',available at:

http://blog.nasscom.in/emerge/2009/08/india-market-and-the-saascloud-computing-landscape/( accessed 19 Jan 2010)

[22] CIOL DQ Channels, 'Is India Ready For Cloud Computing?' available at:

http://dqchannels.ciol.com/content/space/109102902.asp, 2009 (accessed 19 Jan 2010).

[23] CTOFORUM, 'Hosting of IT applications with a third party makes better business sense than managing all of them in-house', available at: http://www.thectoforum.com/content/playing-perfect-host ,2009 ( accessed 4 Feb 2010).

[24] D. Anderson, B.B. Aggarwal, and M. Barnes, "SAP Business ByDesign: Tentative Steps Forward", Springboard Perspective, 2008.

[25] IDC Blog ,'IT Cloud Services Forecast – 2008, 2012: A Key Driver of New Growth', available at:

http://blogs.idc.com/ie/?p=224(accessed 7 Feb 2010).